\documentclass[graybox]{svmult}

\usepackage{helvet}         
\usepackage{courier}        
\usepackage{type1cm}        
\usepackage{makeidx}         
\usepackage{graphicx}        
\usepackage{multicol}        
\usepackage[bottom]{footmisc}

\usepackage{amsmath,amsfonts,amscd,amssymb}
\usepackage{cancel}

\usepackage{latexsym}
\usepackage{overpic}  
\usepackage{epstopdf}
\usepackage{rotating,color}

\graphicspath{{figures/}}

\usepackage{textcomp,xcolor}
\definecolor{MatlabCellColour}{RGB}{250,250,250}
\definecolor{MatPurp}{rgb}{.625,.1406,.9375}
\usepackage{listings}

\lstdefinestyle{customc}{
  belowcaptionskip=.25\baselineskip,
  breaklines=true,
  frame=L,
  xleftmargin=\parindent,
  language=Matlab,
  showstringspaces=false,
  basicstyle=\small\ttfamily,
  keywordstyle=\bfseries\color{white!30!black},
  identifierstyle=\color{blue},  
  commentstyle=\itshape\color{green!60!black},
  stringstyle=\color{MatPurp},
  backgroundcolor=\color{MatlabCellColour}
 }
\newcommand{\bg}{\mathbf{g}}

\newcommand{\bv}{\mathbf{v}}

\newcommand{\bx}{\mathbf{x}}

\newcommand{\bA}{\mathbf{A}}

\newcommand{\bF}{\mathbf{F}}

\newcommand{\bY}{\mathbf{Y}}

\newcommand{\bPhi}{\boldsymbol{\Phi}}

\DeclareMathAlphabet{\mathpzc}{OT1}{pzc}{m}{it}
\usepackage{subeqnar}

\makeindex             

\begin{document}

\title*{Koopman Theory for Partial Differential Equations}
\author{J. Nathan Kutz, Joshua L. Proctor and Steven L. Brunton}
\institute{J. N. Kutz \at Department of Applied Mathematics, University of Washington, Seattle, WA 98195-3925, \email{kutz@uw.edu}
\and J. L. Proctor \at Institute for Disease Modeling, 3150 139th Ave SE, Bellevue, WA 98005, \email{joproctor@intven.com}
\and S. L. Brunton \at Department of Mechanical Engineering, University of Washington, Seattle, WA 98195, \email{sbrunton@uw.edu}}
%
%
\maketitle

\abstract*{PDEs and Koopman.}

\abstract{We consider the application of Koopman theory to nonlinear partial differential equations.  We demonstrate
that the observables chosen for constructing the Koopman operator are critical for enabling an accurate approximation 
to the nonlinear dynamics.  If such observables can be found,
then the dynamic mode decomposition algorithm can be enacted to compute a finite-dimensional approximation  
of the Koopman operator, including its eigenfunctions, eigenvalues and Koopman modes.  Judiciously chosen
observables lead to physically interpretable spatio-temporal features of the complex system under consideration and provide
a connection to manifold learning methods.  We demonstrate the impact of observable selection, including
kernel methods,
and construction of the Koopman operator on two canonical, nonlinear PDEs:  Burgers' equation and the nonlinear Schr\"odinger equation. 
These examples serve to highlight the most pressing and critical challenge of Koopman theory:  a principled way to select appropriate observables. }

\section{Introduction}

Data-driven mathematical methods are increasingly important for characterizing complex systems across the physical, engineering, 
social and biological sciences.  These methods aim to discover and exploit a relatively small subset of the full space where low-dimensional models can be used to describe the evolution of the system.  Thus solutions can often be approximated through dimensionality reduction methods where if $n$ is the dimension of the original high-dimensional system, and $r$ is the dimension of the subspace (or slow-manifold) where the dynamics is embedded, then $r\ll n$.  The reduced order modeling (ROM) community has used this to great effect in applications such as large-scale patterns of atmospheric variability~\cite{majda2012}, turbulent flow control architectures~\cite{Brunton2015amr} and/or spatio-temporal encodings in neurosensory systems~\cite{Ganguli:2012}.
Traditionally, the large-scale dynamics may be embedded in the low-dimensional space using, for instance, the proper orthogonal decomposition (POD) in conjunction with Galerkin projection.   More recently, the dynamic mode decomposition (DMD), and its Koopman generalization, have garnered attention due to the fact that they can (i) discover low-rank spatio-temporal patterns of activity, and (ii) they can embedded the dynamics in the subspace in an equation-free manner, unlike the Galerkin-POD method of ROMs.  In this manuscript, we demonstrate that the Koopman architecture can yield accurate low-dimensional embeddings for nonlinear partial differential equations (PDEs).  Critical to its success is an appropriate choice of observables, which is demonstrated to act as a nonlinear manifold learning method.  We demonstrate the success of the method, and compare it to traditional DMD, on two canonical PDE models:  Burgers' equation and the nonlinear Schr\"odinger equation.

Historically, the DMD method originated in the fluid dynamics community as a principled technique to decompose complex flows into a simple representation based on low-rank, spatio-temporal coherent structures.  Schmid and Sesterhenn~\cite{Schmid2008aps} and Schmid~\cite{Schmid2010jfm} first defined the DMD algorithm and demonstrated its ability to provide physically interpretable insights from high-dimensional fluids data.  The growing success of DMD stems from the fact that it is an \emph{equation-free}, data-driven method capable of providing an accurate decomposition of a complex system into spatio-temporal coherent structures that may be used for diagnostic analysis, short-time future state prediction, and control.  Importantly, Rowley \emph{et al.}~\cite{Rowley2009jfm} showed that DMD is connected to the underlying nonlinear dynamics through Koopman operator theory~\cite{Koopman1931pnas} and is readily interpretable using standard dynamical systems techniques~\cite{Mezic2004,Mezic2005nd,budivsic2012applied,Mezic2013arfm}.  Specifically, the DMD algorithm is a manifestation of Koopman theory when the observable functions are the identify or a linear transformations of the underlying state space.  Thus DMD is a principled, algorithmic architecture allowing for an explicit approximation of the Koopman operator.  For more details, there are numerous detailed references~\cite{Rowley2009jfm,Tu2014jcd,Kutz2016book}.  

The approximation of the Koopman operator via DMD is critically important for enabling evaluation of the operator from data.  Indeed, it transforms Koopman theory from an abstract mathematical conception to a readily tractable computation.   It also highlights the important
role played by observables and their associated evolution manifolds.  In particular, nonlinear PDEs can be thought to evolve on manifolds which are often difficult to characterize and are rarely known analytically.  A correct choice of observables can, in some cases, {\em linearize} the nonlinear manifold.  For instance, the nonlinear evolution governed by Burgers' PDE equation can be linearized by the Cole-Hopf transformation, thus providing a linear manifold which can trivially describe the evolution dynamics.  Such exact solutions to nonlinear PDEs are extremely rare and do not often exist in practice, with the inverse scattering transform (IST) for Korteweg-deVries, nonlinear Schr\"odinger and other integrable PDEs being the notable exceptions.   Regardless, judiciously chosen observables can help transform a PDE evolving on a strongly nonlinear manifold to a weakly nonlinear manifold, enabling a more accurate and  broader range of applicability of the Koopman approximation.

\section{Koopman Theory, Observables, and Dynamic Mode Decomposition}

The original work of Koopman in 1931~\cite{Koopman1931pnas} considered Hamiltonian systems and formulated the Koopman operator as a discrete-time mapping.  We generalize the Koopman operator definition for a continuous time system.  \\

\noindent {\bf Definition:  Koopman\index{Koopman theory} Operator}: {\em Consider a continuous-time dynamical system\index{dynamical system}
\begin{equation}
  \frac{d{\bf x}}{dt} = \bf{f}({\bf x}) \, ,
  \label{eq:Ukoop}
\end{equation}
where ${\bf x}\in\mathcal{M}$ is the state on a smooth $n$-dimensional manifold $\mathcal{M}$.  The Koopman operator ${\cal K}$ is an infinite-dimensional linear operator that acts on all observable functions $g:\mathcal{M}\rightarrow\mathbb{C}$ so that:}
\begin{equation}
  {\cal K} {g}({\bf x})  = { g}\left( {\bf f}({\bf x}) \right) \, . \label{eq:koop}
\end{equation}
In the following year, Koopman and von Neumann extended these results to dynamical systems with continuous spectra~\cite{Koopman1932pnas}.  Critical to implementing this definition numerically is understanding how to choose a finite set of 
observables ${g}({\bf x})$.  This remains an open challenge today and will be addressed in our PDE examples.

By construction, the Koopman operator is a {\em linear}, infinite-dimensional operator that acts on the Hilbert space $\mathcal{H}$ of \emph{all} scalar measurement functions $g$.  The Koopman operator acts on functions of the state space of the dynamical system,  trading nonlinear finite-dimensional dynamics for linear infinite-dimensional dynamics. It can be further generalized to map infinite-dimensional nonlinear dynamics to infinite-dimensional linear dynamics by appropriate choice of observables.  In practice, the computation of the Koopman operator will require a finite-dimensional representation. The advantage of the Koopman representation is compelling:  linear problems can be solved using standard linear operator theory and spectral decompositions.   With such methods the infinite dimensional representation is handled by considering a sufficiently large, but finite, sum of modes to approximate the Koopman spectral solution.  It should be noted that the definition (\ref{eq:koop}) can be alternatively represented by a composition of the observables with the nonlinear evolution:  ${\cal K} g =  g \circ {\bf f}$.  

The Koopman operator may also be defined for discrete-time dynamical systems, which are more general than continuous-time systems.  In fact, the dynamical system in \eqref{eq:Ukoop} will induce a discrete-time dynamical system given by the flow map $\bF_t:\mathcal{M}\rightarrow\mathcal{M}$, which maps the state $\bx(t_0)$ to a future time $\bx(t_0+t)$:
\begin{eqnarray}
\bF_t(\bx(t_0)) = \bx(t_0+t) = \bx(t_0) + \int_{t_0}^{t_0+t}\bf{f}(\bx(\tau))\,d\tau\,.
\end{eqnarray}
This induces the discrete-time dynamical system
\begin{eqnarray}
\bx_{k+1} = \bF_t(\bx_k),\label{Eq:Dynamics}
\end{eqnarray}
where $\bx_k=\bx(kt)$.  The analogous discrete-time Koopman operator is given by $\mathcal{K}_t$ such that $\mathcal{K}_tg = g\circ\bF_t$.  
Thus, the Koopman operator sets up a discrete-time dynamical system on the observable function $g$:
\begin{eqnarray}
\mathcal{K}_tg(\bx_k) = g(\bF_t(\bx_k)) = g(\bx_{k+1}).
\end{eqnarray}

If an appropriate Koopman operator can be constructed, then linear operator theory provides the spectral decomposition required to represent the dynamical solutions of interest.  Specifically, the eigenfunctions and eigenvalues of the Koopman operator ${\cal K}$ give a complete characterization of the dynamics.  We consider the eigenvalue problem
\begin{subeqnarray}
  &&  {\cal K} \varphi_k = \lambda_k \varphi_k.  \label{eq:koopeig}
\end{subeqnarray}
The functions $\varphi_k(\bx)$ are Koopman eigenfunctions, and they define a set of intrinsic measurement coordinates, on which it is possible to advance these measurements with a \emph{linear} dynamical system.  The low-dimensional embedding of the dynamics are ultimately extracted from the Koopman eigenfunctions.  More precisely, a reduced-order linear model can be constructed by a rank-$r$ truncation of the dominant eigenfunctions $\varphi_k$.

A vector of observables $\bg$, which is in our new {\em measurement space}, may be expressed in terms of Koopman eigenfunctions $\varphi$ as
\begin{eqnarray}
\bg(\bx) = \begin{bmatrix} g_1(\bx)\\ g_2(\bx) \\ \vdots\\  g_p(\bx)\end{bmatrix} = \sum_{k=1}^{\infty}\varphi_k(\bx)\bv_k,\label{eq:koopexpand}
\end{eqnarray}
where $\bv_k$ is the $k$-th  Koopman mode associated with the $k$-th  Koopman eigenfunction $\varphi_k$, i.e. it is the weighting of
each observable on the eigenfunction.   In the original theory~\cite{Koopman1931pnas}, Koopman considered Hamiltonian flows that are measure preserving, so that the Koopman operator is unitary.  In this case, the eigenfunctions are all orthonormal, and \eqref{eq:koopexpand} may be written explicitly as:
\begin{eqnarray}
\bg(\bx) = \sum_{k=1}^{\infty}\varphi_k(\bx)\begin{bmatrix} \langle \varphi_k,g_1\rangle\\ \langle \varphi_k,g_2\rangle\\ \vdots \\ \langle \varphi_k,g_p\rangle \end{bmatrix} = \sum_{k=1}^{\infty}\varphi_k(\bx)\bv_k.
\end{eqnarray}
The dynamic mode decomposition algorithm is used to compute an approximation to the Koopman eigenvalues $\lambda_k$ and modes $\bv_k$.  

The nonlinear dynamical system defined by ${\bf f}$ in (\ref{eq:Ukoop}) and the infinite-dimensional, linear dynamics defined by 
${\cal K}$ in (\ref{eq:koop}) are equivalent representations of a dynamical system.   One can either evolve the system in the original state space (\ref{eq:Ukoop}), requiring computational effort since it is nonlinear, or one can instead evolve using (\ref{eq:koop}) and (\ref{eq:koopexpand}) so that the time dynamics are trivially computed
\begin{equation}
    {\cal K} {\bf g}({\bf x}) = {\cal K} \sum_{k=1}^\infty  \varphi_k ({\bf x}) {\bf v}_k   = \sum_{k=1}^\infty   {\cal K} \varphi_k ({\bf x}){\bf v}_k 
    = \sum_{k=1}^\infty   \lambda_k \varphi_k ({\bf x}) {\bf v}_k  .  
\end{equation}
Thus future solutions can be computed by simple multiplication with the Koopman eigenvalue\index{Koopman eigenvalues}.
Such a mathematical strategy for evolving nonlinear dynamical systems would seem always to be advantageous.  However, it remains an open challenge how to systematically link the observables\index{observables} ${\bf g}$ and the associated  Koopman mode expansion to the original evolution defined by ${\bf f}$.  For a limited class of nonlinear dynamics, this can be done explicitly~\cite{Brunton2016plosone}.

\begin{figure}[t]
\vspace*{-.9in}
\hspace*{-.7in}
\begin{overpic}
[width=1.25\textwidth]{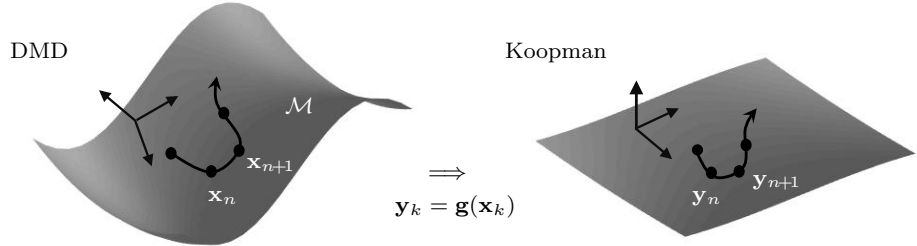}
\put(28,37){\color{white}{${\bf x}_n$}}
\put(31.5,40){\color{white}{${\bf x}_{n\!+\!1}$}}
\put(72,37){\color{white}{${\bf y}_n$}}
\put(77.5,38.5){\color{white}{${\bf y}_{n\!+\!1}$}}
\put(45,36){${\bf y}_k={\bf g}({\bf x}_k)$}
\put(35,45){\color{white}{$\mathcal{M}$}}
\put(48,39){$\Longrightarrow$}
\put(10,50){DMD}
\put(55,50){Koopman}
\end{overpic}
\vspace*{-1.9in}
\caption{The left panel illustrates the nonlinear manifold on which the dynamical system defined by ${\bf F}_t$ in (\ref{Eq:Dynamics}) generates
a solution.  DMD approximates the evolution on this manifold by a least-square fit linear dynamical system.  In contrast, the selection of appropriate observables $g({\bf x})$ define a Koopman operator that helps {\em linearize} the manifold so that a least-square fit linear dynamical system provides a much better approximation to the system (right panel).  
\label{fig:koop1}}
\end{figure}

Figure~\ref{fig:koop1} illustrates the underlying concept in the Koopman approach.   A dynamical system consisting of snapshots ${\bf x}_k$ evolves according to the nonlinear dynamical system defined by ${\bf F}_t$ in (\ref{Eq:Dynamics}).  In the state space, the nonlinearity generates a nonlinear manifold in which the data is embedded.  The DMD approximation produces a least-square fit {\em linear dynamical system} approximating the flow map and the low-dimensional embedding (left panel of Fig.~\ref{fig:koop1}).  Koopman theory ideally defines an operator that attempts to {\em linearize} the space in which the data is embedded.   The Koopman operator then produces a linear flow map and low-dimensional embedding that approximates the full nonlinear dynamics (right panel of Fig.~\ref{fig:koop1}).

\section{The DMD and Koopman algorithms}
\label{sec:dmd}

The DMD algorithm underlies the computation of the Koopman eigenvalues and modes directly from data.  Its effectiveness depends sensitively on the choice of observables.   Rowley \emph{et al.}~\cite{Rowley2009jfm} showed that DMD approximates the Koopman operator for the set of
observables $\mathbf{g}({\bf x})={\bf x}$.  We will use this fact in constructing a DMD algorithm for observables of ${\bf x}$ instead of the state variable itself. To start, we use the following definition of the DMD decomposition~\cite{Tu2014jcd}:\\

\noindent {\bf Definition:  Dynamic Mode Decomposition} (Tu \emph{et al.} 2014~\cite{Tu2014jcd}): {\em Suppose we have
a dynamical system\index{dynamical system} (\ref{Eq:Dynamics}) and two sets of data 
%
\begin{subeqnarray}
 && {\bf X} = \begin{bmatrix}
\vline & \vline & & \vline \\
\bx_1 & \bx_2 & \cdots & \bx_{m}\\
\vline & \vline & & \vline
\end{bmatrix} \\
 && {\bf X}' = \begin{bmatrix}
\vline & \vline & & \vline \\
\bx'_1 & \bx'_2 & \cdots & \bx'_{m}\\
\vline & \vline & & \vline
\end{bmatrix}
\end{subeqnarray}
%
with ${\bf x}_k$ an initial condition to (\ref{Eq:Dynamics}) and ${\bf x}'_k$ it corresponding output after some prescribed evolution time $\Delta t$ with there being $m$ initial conditions considered.  The DMD modes are eigenvectors of 
\begin{equation}
  {\bf A}_{\bf X} = {\bf X}' {\bf X}^\dag
  \label{eq:newDMD}
\end{equation}
where $\dag$ denotes the Moore-Penrose pseudoinverse.}\\

The above definition provides a computational method for evaluating the Koopman operator for a linear observable. 
In practice, three practical constraints must be considered:  (i) We have data ${\bf X}$ and ${\bf X}'$, but
we do not necessarily know ${\bf F}_t(\cdot)$,  (ii) We will have to make a finite-dimensional approximation to the
infinite-dimensional Koopman operator ${\cal K}$, and (iii) We will have to judiciously select the observables
$g({\bf x})$ in order to have confidence that that Koopman operator will approximate the nonlinear dynamics
of ${\bf F}_t(\cdot)$.  Points (i) and (ii) go naturally together.  Specifically, the number of measurements in
each column of ${\bf X}$ and ${\bf X}'$ are $n$, while the number of total columns (time measurements) is $m$.
Thus finite-dimensionality is imposed simply from the data collection limitations.  The dimension can
be increased with a large set of observables, or it can be decreased via a low-rank truncation during the DMD
process.  The observables are more difficult to deal with in a principled way.  Indeed, a good choice of 
observables can make the method extremely effective, but it would also require expert knowledge of the
system at hand~\cite{Brunton2016plosone}.  This will be discussed further in the examples.

The following gives a practical demonstration of how to use the data, the DMD algorithm, and the observables to
produce a Koopman operator and a future state prediction of the nonlinear evolution (\ref{eq:Ukoop}).
The Koopman algorithm simply applies DMD on the space of observables.\\

\begin{enumerate}

\item From the data matrices ${\bf X}$ and ${\bf X}'$, create
the data matrices of observables ${\bf Y}$ and ${\bf Y}'$:
%
\begin{subeqnarray}
 && {\bf Y} = \begin{bmatrix}
\vline & \vline & & \vline \\
{\bf g}(\bx_1) & {\bf g}(\bx_2) & \cdots & {\bf g}(\bx_{m-1})\\
\vline & \vline & & \vline
\end{bmatrix} \\
 && {\bf Y}' = \begin{bmatrix}
\vline & \vline & & \vline \\
{\bf g}(\bx'_1) & {\bf g}(\bx'_2) & \cdots & {\bf g}( \bx'_{m-1})\\
\vline & \vline & & \vline
\end{bmatrix}
\end{subeqnarray}
%
where each column is given by ${\bf y}_k={\bf g}({\bf x}_k)$ or ${\bf y}'_k={\bf g}({\bf x}'_k)$

\item Perform the DMD algorithm on the pair ${\bf Y}$ and ${\bf Y}'$ to compute 
\begin{equation}
{\bf A}_{\bf Y}={\bf Y}'{\bf Y}^\dag
\end{equation}
along with the low-rank counterpart $\tilde{\bf A}_{\bf Y}$ obtained by projection onto a truncated POD subspace.  The eigenvalues and eigenvectors of $\bA_\bY$ may approximate Koopman eigenvalues and modes if the observables are well chosen.  

\item DMD can be used to compute the augmented modes $\bPhi_{\bY}$, which may approximate the Koompan modes, by
\begin{equation}
\bPhi_{\bY} = {\bf Y}' {\bf V} {\bf \Sigma}^{-1} {\bf W}
\end{equation}
where ${\bf W}$ comes from the eigenvalue problem $\tilde{\bf A}_{\bf Y} {\bf W}={\bf W} {\bf \Lambda}$ and
${{\bf Y}={\bf U}{\bf \Sigma}{\bf V}^*}$.  Note that an $r$-rank truncation of the SVD is performed at this stage.

\item The future state in the space of observables is given by the linear evolution 
\begin{equation}
  {\bf y}(t) = \bPhi_{\bY} \, \mbox{diag}(\exp(\boldsymbol{\omega} t)) \,{\bf b}
\end{equation}
where ${\bf b}=\bPhi_{\bY}^\dag {\bf y}_1$ is determined by projecting back
to the initial data observable.  The continuous-time eigenvalues $\boldsymbol{\omega}$ are obtained from the discrete-time eigenvalues $\lambda_k$ (i.e., diagonal elements of the matrix ${\bf \Lambda}$) where $\omega_k = \ln (\lambda_k)/\Delta t$.

5.  Transform from observables to state space  
\begin{equation}
{\bf y}_k= {\bf g}({\bf x}_k) \,\,\,\,\, \rightarrow \,\,\,\,\, {\bf x}_k={\bf g}^{-1} ({\bf y}_k) .
\end{equation}
This last step is trivial if one of the observables selected to comprise ${\bf g}({\bf x}_k)$ is
the state variable ${\bf x}_k$ itself.  If only nonlinear observables of ${\bf x}_k$ are chosen,
then the inversion process can be difficult.

\end{enumerate}

This process shows that the DMD algorithm is closely related to the Koopman operator.  Indeed, it
is the foundational piece for practical evaluation of the finite-dimensional Koopman operator.
It is stressed once again here:  selection of appropriate observables is critical for the algorithm
to generate good reconstructions and approximations to the future state.
We can also now introduce the following theorem~\cite{Rowley2009jfm,Tu2014jcd,Rowley2014ipam,Williams2015jnls}.\\

\noindent {\bf Theorem:  Koopman and Dynamic Mode Decomposition}: {\em 
Let $\varphi_k$ be an eigenfunction of ${\cal K}$ with eigenvalue $\lambda_k$, and suppose
$\varphi_k\in \mbox{span} \{ g_j \}$, so that
\begin{equation}
   \varphi_k ({\bf x})= w_1 g_1({\bf x}) + w_2 g_2({\bf x}) + \cdots + w_p g_p({\bf x}) = {\bf w} \cdot {\bf g}
\end{equation}
for some ${\bf w}=[w_1 \,\,\, w_2 \,\,\, \cdots \,\,\, w_p]^T\in \mathbb{C}^p$. If ${\bf w}\in R({\bf Y})$, where $R$ is the range, then
${\bf w}$ is a left eigenvector of ${\bf A}_{\bf Y}$ with eigenvalue $\lambda_k$ so that 
$\tilde{\bf w}^* {\bf A}_{\bf Y}=\lambda_k \tilde{\bf w}^*$. }\\

Thus the Koopman eigenvalues are the DMD eigenvalues provided
(i) the set of observables is sufficiently large so that $\varphi_k({\bf x})\in\mbox{span} \{ g_j \}$ and
(ii) the data is sufficiently {\em rich} so that  ${\bf w} \in R({\bf X})$.
This directly shows that the choice of observables is critical in allowing one to connect DMD
theory to Koopman spectral analysis.  If this can be done, then one can simply take data snapshots
of a finite-dimensional nonlinear dynamical system in time and re-parameterize it as a linear system in the observable coordinates, which is amenable to a simple eigenfunction (spectral) decomposition.
This representation diagonalizes the dynamics and shows that the time evolution of each eigenfunction corresponds to multiplication by its corresponding eigenvalue.

\section{Koopman Observables and Kernel Methods}

The effectiveness of Koopman theory hinges on one thing:  selecting appropriate observables.  
Once observables are selected, the previous section defines a DMD-based algorithm for computing
the Koopman operator whose spectral decomposition completely characterizes the approximation.
In the machine learning literature, observables  are often thought of as {\em features}, and we will
build upon this concept to generate appropriate observables.  An important practical consideration 
becomes the computational cost in generating the DMD approximation as the number of rows in the matrices ${\bf Y}$ and ${\bf Y}'$ get progressively larger with each additional observable.

In the absence of expert-in-the-loop knowledge of the dynamical system, one might
consider, for instance, the support vector machine (SVM) literature and associated
kernel methods~\cite{svm1,svm2,svm3,svm4} for feature selection (observables).   The SVM architecture suggests a number
of techniques for constructing the feature space ${\bf g}({\bf x})$, with a common choice being
the set of polynomials such that 
\begin{equation}
g_j (x)= \left\{ x, x^2, x^3, x^4, \hdots , x^p \right\} \, .
\label{eq:ext_obs}
\end{equation}
Using a large number of polynomials can generate an extremely large vector of observables for
each snapshot in time.  This is closely related to the Carleman linearization technique in dynamical systems~\cite{steeb1980non,kowalski1991nonlinear,banks1992infinite}.
Alternatively, kernel methods have found a high degree of success using (i) radial basis functions, typically
for problems defined on irregular domains, (ii) Hermite polynomials for problems defined on $\mathbb{R}^n$,
and (iii) discontinuous spectral elements for large problems with block diagonal structures.
Regardless of the specific choice of feature space, the goal is to choose a sufficiently rich and
diverse set of observables that allow an accurate approximation of the Koopman operator ${\cal K}$.
Instead of choosing the correct observables, one then simply chooses a large set of candidate
observables with the expectation that a sufficiently diverse set will include enough features for
an accurate reconstruction of the Koopman modes, eigenfunctions and eigenvalues, which intrinsically characterize the nonlinear dynamical system.

Williams \emph{et al.}~\cite{Williams2015jnls,Williams2014arxivA} have recently capitalized on the ideas
of machine learning by implementing the so-called extended DMD and kernel DMD method on extended observables (\ref{eq:ext_obs})
within the DMD architecture.  Moreover, they have developed an efficient way to compute 
$\tilde{\bf A}_{\bf Y}$ even for a large observable space.  The kernel DMD method is the most relevant in practice as the number of
observables (features) can rapidly grow so as to make $n$ extremely high-dimensional.
In the context of the Koopman operator, the kernel trick~\cite{svm1,svm2,svm3,svm4} will define a function $h({\bf x},{\bf x}')$ that can be related to the observables $g_j({\bf x})$ used for constructing ${\bf Y}$ and ${\bf Y}'$.  Consider the simple example of a quadratic polynomial kernel
\begin{equation}
  h ({\bf x},{\bf x}') = 
    \left(  1+ {{\bf x}}^T {\bf x}' \right)^2
    \label{eq:kernel1}
\end{equation}
where ${\bf x}$ and ${\bf x}'$ are data points in $\mathbb{R}^{2}$.  
When expanded, the kernel function takes the form
\begin{eqnarray}
   h ({\bf x},{\bf x}') &=& \left( 1 + x_1 x_1' + x_2 x_2' \right)^2 \nonumber \\
   &=&  \left( 1 + 2 x_1 x_1' + 2 x_2 x_2' + 2 x_1 x_2 x_1' x_2'  + x_1^2 {x_1'}^2 + x_2^2 {x_2'}^2 \right) \nonumber \\
   &=& {\bf Y}^T ({\bf x}')  {\bf Y} ( {\bf x} )   \label{eq:kernel3}
\end{eqnarray}
where  ${\bf Y} ({\bf x}) = \left[  \sqrt{2} x_1 \,\, \sqrt{2} x_2 \,\, \sqrt{2} x_1 x_2 \,\,
{x_1}^2 \,\, {x_2}^2 \right]^T $.
Note that for this case, both the Koopman observables and the 
kernel function (\ref{eq:kernel1}) are equivalent representations that are paired together through
the expansion (\ref{eq:kernel3}).  The so-called kernel trick posits that (\ref{eq:kernel1}) is
a significantly more efficient representation of the polynomial variables that emerge from
the expansion (\ref{eq:kernel3}).  Instead of defining the Koopman
observables $g_i({\bf x})$, we instead
define the kernel function (\ref{eq:kernel1}) as it provides a compact representation of
the feature space and an implicit computation of the inner products required for the Koopman operator. 

The computational advantages of the kernel trick are considerable.  For example, 
a polynomial kernel of degree $p$ acting on data vectors ${\bf x}$ and ${\bf x}'$  in $\mathbb{R}^{n}$ is
given by
\begin{equation}
  h ({\bf x},{\bf x}') = 
    \left(  1+ {{\bf x}}^T {\bf x}' \right)^p
    \label{eq:kernelp}
\end{equation}
which requires a single computation of the inner product $\alpha={{\bf x}}^T {\bf x}'$.  This requires ${\mathcal O} (n)$ and produces
$f ({\bf x},{\bf x}') = (1+\alpha)^p$ where $\alpha$ is a constant.  The resulting computational cost for this $p$th degree polynomial kernel 
remains ${\mathcal O} (n)$.
In contrast, the equivalent observable space using ${\bf Y}$ requires construction of a vector of length $O(n^2)$ 
taking the form
\begin{equation}
{\bf Y} ({\bf x}) = \begin{bmatrix} 1 & x_1 &  \cdots & x_n & x_1^2 & x_1x_2 & \cdots & x_n^2 & \cdots & x_1^p & \cdots & x_n^p \end{bmatrix}^T \,.
\label{eq:kernelhuge}
\end{equation}
Computing the inner product  ${\bf Y}^T ({\bf x}')  {\bf Y} ( {\bf x} )$ is
a significantly larger computation than the kernel form (\ref{eq:kernelp}).  Thus the kernel trick enables an advantageous
representation of the various polynomial terms and circumvents the formation and computation associated with (\ref{eq:kernelhuge}).

The choice of kernel is important and in practice, is not robust for Koopman methods.  Some standard choices are often used, including the three most common kernels of SVM-based data methods:
\begin{subeqnarray}
  &\hspace*{-.5in} \text{polynomial kernel (degree $p$)}  &  h ({\bf x},{\bf x}')= \left( a + {\bf x}^T{\bf x}' \right)^p \\
  &\hspace*{-.5in} \text{radial basis functions\index{radial basis functions}} &   h ({\bf x},{\bf x}')= \exp \left( -a | {\bf x} - {\bf x}' |^2 \right) \\
  &\hspace*{-.5in} \text{sigmoid kernel\index{sigmoid kernel}}  &  h ({\bf x},{\bf x}')= \tanh \left(  {\bf x}^T{\bf x}' + a \right) .
  \label{eq:kernel4}
\end{subeqnarray}
The advantage of the kernel trick is quite clear, providing a compact representation of a very large feature space.  
For the polynomial kernel, for instance,
a 20th-degree polynomial $(p=20)$ using (\ref{eq:kernel4}a) is trivial and does not compute all the inner products directly.  In contrast, using our
standard Koopman observables $g({\bf x}_j)$ would require one to explicitly write out all the terms
generated from a 20th-degree polynomial on an $n$-dimensional data set, which is computationally intractable for even moderately large $n$.  
The tuning parameter $a$ must be carefully chosen in practice for reasonable results.

In practice, the observables for ${\bf Y}$ are implicitly embedded in the kernel $h({\bf x},{\bf x}')$.  Specifically,
we consider the observable matrix elements defined by
 \begin{equation}
  {\bf Y}^T {\bf Y}' (j,k) = h({\bf x}_j,{\bf x}'_k)
  \label{eq:feature1}
\end{equation}
where the $(j,k)$ denotes the $j$th row and $k$th column of the correlation matrix, and the ${\bf x}_j$ and
${\bf x}'_k$ are the $j$th and $k$th columns of data.  The kernel
DMD formulation still requires the computation of the matrices ${\bf V}$ and
${\bf  \Sigma}$ which can be produced from
$ {\bf Y}^* {\bf Y} {\bf V} = {\Sigma}^2 {\bf V} $.
As before, the matrix elements of ${\bf Y}^* {\bf Y}$ are computed from $ {\bf Y}^* {\bf Y} (j,k) = h({\bf x}_j,{\bf x}_k)$ .
Thus all the required inner products are computed by projecting 
directly to the feature space.   Note that if the linear kernel function $h({\bf x},{\bf x})={\bf x}^T {\bf y}$ is chosen,
the kernel DMD reduces to the standard DMD algorithm.

\section{Application to PDEs}

To demonstrate the Koopman operator concepts, we apply the methodology to two illustrative and canonical
PDEs:  Burgers' equation and the nonlinear Schr\"odinger equation.  With these two examples, we can (i) illustrate
a scenario where the Koopman operator can exactly (analytically) linearize a dynamical system, (ii) demonstrate
how to judiciously select observables, and (iii) show that kernel methods are highly sensitive as an observable selection
technique.

\subsection{Burgers' Equation}

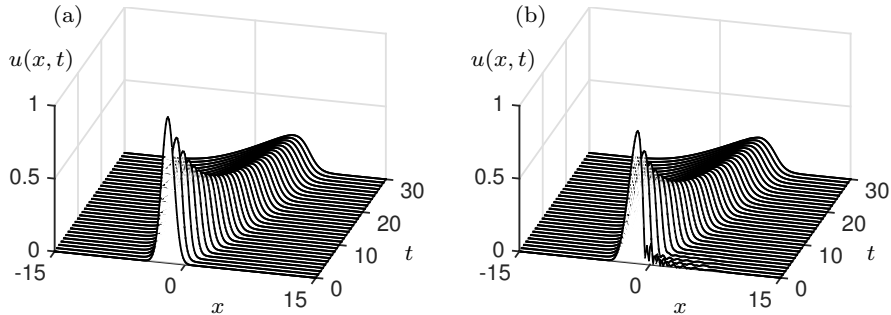
\begin{figure}[t]
\vspace*{.3in}
\hspace*{-.0in}
\begin{overpic}
[width=1.0\textwidth]{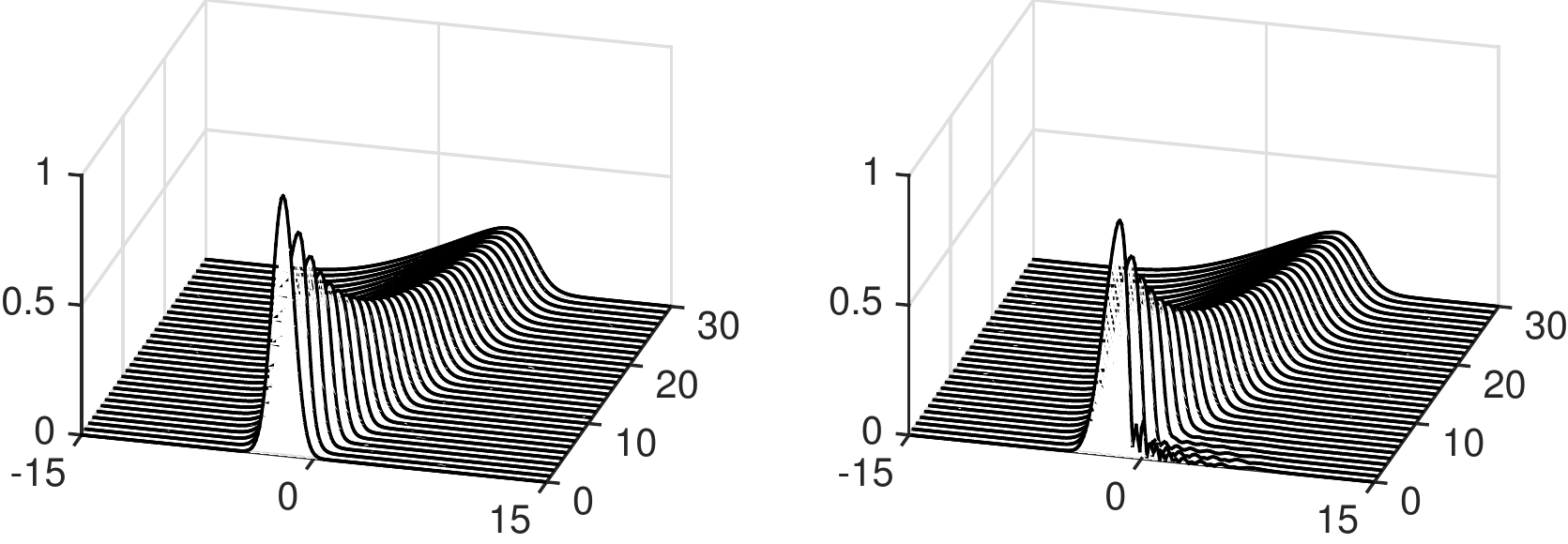}
\put(0,28){$u(x,t)$}
\put(23,0){$x$}
\put(45,6){$t$}
\put(52.5,28){$u(x,t)$}
\put(75.5,0){$x$}
\put(97.5,6){$t$}
\put(5,33){(a)}
\put(57.5,33){(b)}
\end{overpic}
\vspace*{-0.1in}
\caption{(a) Evolution dynamics of Burgers' equation with initial condition $u(x,0)=\exp(-(x+2)^2)$.  (b)  Fifteen mode DMD approximation 
of the Burgers' evolution. 
\label{fig:burg1}}
\end{figure}

To demonstrate the construction of a specific and exact Koopman operator, we consider the canonical nonlinear PDE:   Burgers' equation with 
diffusive regularization.  The evolution, as illustrated in Fig.~\ref{fig:burg1}(a), is governed by diffusion with a nonlinear advection~\cite{burgers}:
\begin{equation}
  u_t + u u_x - \epsilon u_{xx} = 0  \,\,\,\,\,\,\,\, \epsilon>0, \,\,  x\in[-\infty,\infty].
  \label{burgers}
\end{equation}
When $\epsilon=0$, the evolution can lead to shock formation in finite time.  The presence of the diffusion term regularizes the PDE, ensuring continuous solutions for all time. 

Burgers' equation is one of the few nonlinear PDEs whose analytic solution form can be derived.  In independent, seminal contributions, 
Hopf~\cite{hopf50} and Cole~\cite{cole51} derived a transformation that linearizes the PDE.  The Cole-Hopf transformation
is defined as follows
\begin{equation}
  u=-2\epsilon v_x/v \, .
  \label{colehopf}
\end{equation}
The transformation to the new variable $v(x,t)$ replaces the nonlinear PDE (\ref{burgers}) with the linear, diffusion equation
\begin{equation}
  v_t = \epsilon v_{xx}
\end{equation}
where it is noted that $\epsilon>0$ in (\ref{burgers}) in order to produce a well-posed PDE.

The diffusion equation can be easily solved using Fourier transforms.  Fourier transforming in $x$ gives
the ODE system
\begin{equation}
  \hat{v}_t = - \epsilon k^2 \hat{v}
\end{equation}
where $\hat{v}=\hat{v}(k,t)$ denotes the Fourier transform of $v(x,t)$ and $k$ is the wavenumber. 
The solution in the Fourier domain is easily found to be
\begin{equation}
  \hat{v}=\hat{v}_0 \exp( -\epsilon k^2 t )
  \label{bkoop}
\end{equation}
where $\hat{v}_0=\hat{v}(k,0)$ is the Fourier transform of the initial condition $v(x,0)$.

\begin{figure}[t]
\vspace*{.2in}
\hspace*{.20in}
\begin{overpic}
[width=0.9\textwidth]{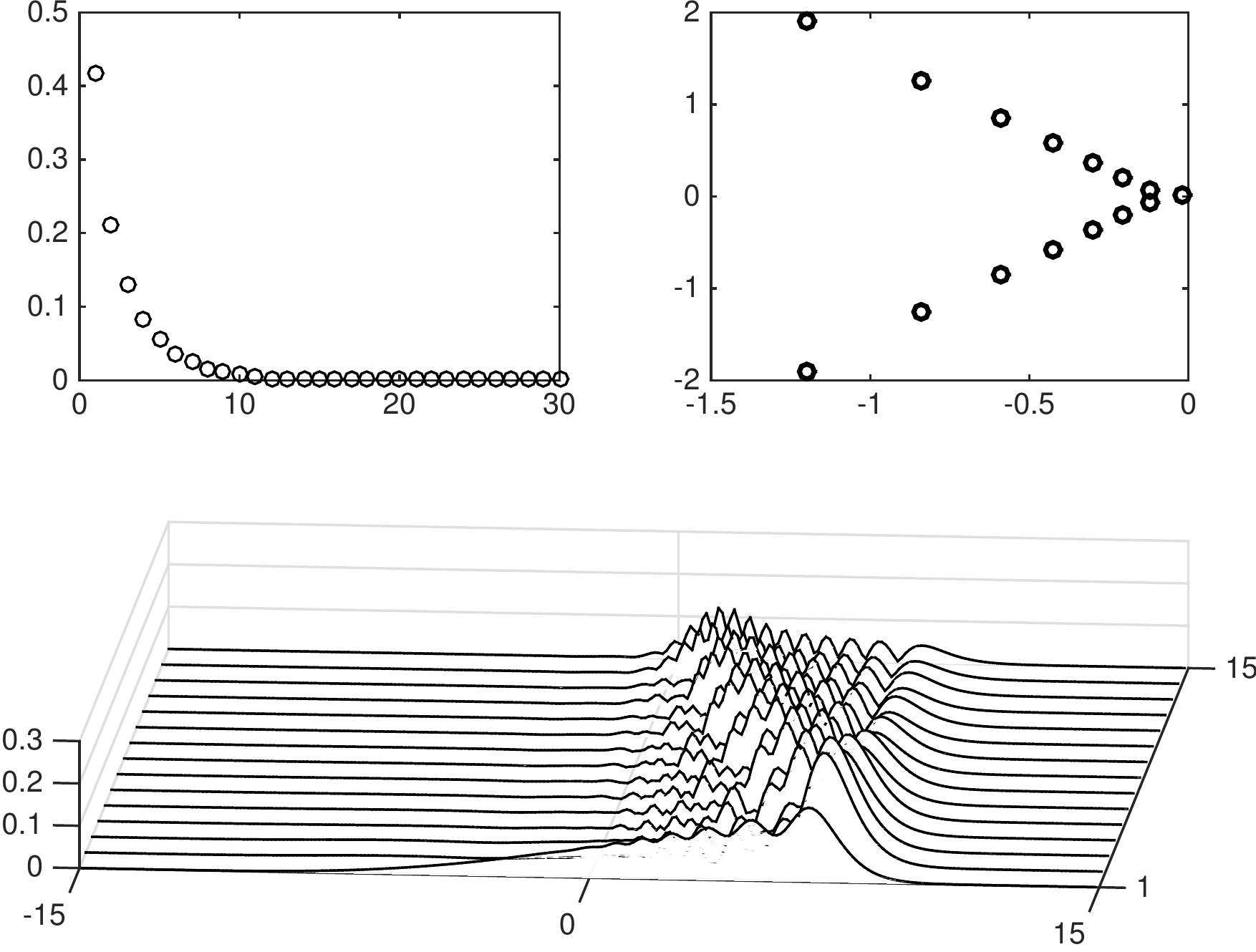}
\put(10,68){(a)}
\put(60,68){(b)}
\put(10,30){(c)}
\put(46.5,69.5){$\Re\{\omega_k\}$}
\put(85,38){$\Im\{\omega_k\}$}
\put(-5,69.5){$\sigma_k$}
\put(20,38){{mode number}}
\put(60,0){{$x$}}
\put(95,5){\rotatebox{70}{mode number}}
\put(-5,20){$\varphi_k$}
\end{overpic}
\vspace*{0.05in}
\caption{DMD of the Burgers' equation.  (a) The singular value spectrum demonstrates that a rank $r=15$ truncation
should be adequate to capture the dynamics of the front propagation in Fig.~\ref{fig:burg1}.  (b)  The eigen-decomposition in the
DMD algorithm produces a DMD spectra whose eigenvalues are decaying.  (c)  The DMD modes used for reconstructing the solution
in Fig.~\ref{fig:burg1} ordered according the smallest to largest (in magnitude) eigenvalues.  The first mode is like a {\em background} mode since the eigenvalue is almost zero.
\label{fig:burg2}}
\end{figure}
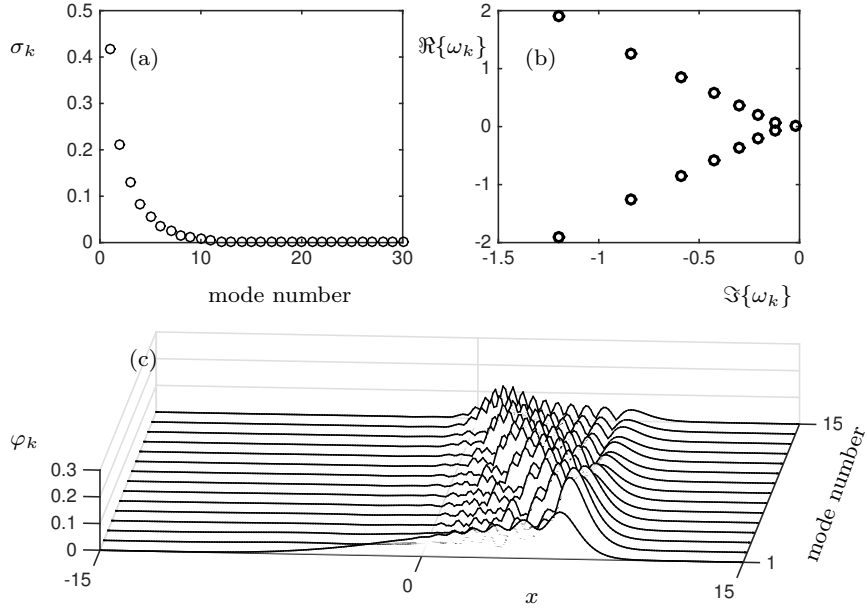

To construct the Koopman operator, we can then combine the transform to
the variable $v(x,t)$ from (\ref{colehopf})
\begin{equation}
  v(x,t)= \exp \left[ -\frac{\int_{-\infty}^{x} u(\xi,t) d\xi}{2\epsilon} \right]
\end{equation}
with the Fourier transform to define the observables
\begin{equation}
    g(u) = \hat{v} \, .
    \label{eq:burg_obs}
\end{equation}
The Koopman operator is then constructed from (\ref{bkoop}) so that
\begin{equation}
   {\cal K} = \exp( -\epsilon k^2 t )  \, .
   \label{eq:burg_koo}
\end{equation}
This is one of the rare instances where an explicit expression for the Koopman operator and the 
observables can be constructed analytically.   The inverse scattering transform for other canonical
PDEs, KdV and NLS,  also can lead to an explicit expression for the Koopman operator, but the
scattering transform and its inversion are much more difficult to construct in practice.

To make comparison between Koopman theory and DMD, we consider the DMD method applied
to governing equation (\ref{burgers}).  Applying the algorithm of Sec.~\ref{sec:dmd} to the observables $g({\bf x})={\bf x}$
gives the DMD approximation to the Burgers' dynamics as shown in Fig.~\ref{fig:burg1}(b).  For this simulation, data
snapshots where collected at intervals of $\Delta t=1$ for the time range $t\in[0,30]$.  The singular value decay for the
dynamics is shown in Fig.~\ref{fig:burg2}(a), suggesting that a rank $r=15$ truncation is appropriate.  The DMD spectra
and DMD modes are illustrated in Fig.~\ref{fig:burg2}(b) and (c) respectively.  Thus using $u(x,t)$ directly as an observable
produces a low-rank model with fifteen modes.  In contrast, by working with the observable (\ref{eq:burg_obs}), the Koopman
operator can be trivially computed (\ref{eq:burg_koo}) and the dynamics analytically produced without need of approximation.
In this case, the Koopman operator exactly linearizes the dynamics.  This is the ideal which is hoped for, but rarely achieved
with nonlinear PDEs (or nonlinear dynamical systems in general).

\subsection{Nonlinear Schr\"odinger Equation}

The example of Burgers' equation was easy to quantify and understand since the Cole-Hopf transformation was discovered
nearly seven decades ago.  Thus the observables chosen were easily motivated from knowledge of the analytic solution.
Unfortunately, it is rarely the case that such linearizing transformations are known.  In our second example, we consider the
Koopman operator applied to a second canonical nonlinear PDE:  the nonlinear Schr\"odinger equation
\begin{equation}
  i  u_t + \frac{1}{2} u_{xx} + |u|^2 u =0
  \label{eq:nls_koop}
\end{equation}
where $u(x,t)$ is a function of space and time modeling slowly-varying optical fields or deep water waves, for instance. 
Discretizing in the spatial variable $x$, we can Fourier transform the solution in space and use a 
standard time-stepping algorithm, such as a fourth-order Runge-Kutta, to integrate the solution forward in
time.  

\begin{figure}[t]
\vspace*{.1in}
\hspace*{-.20in}
\begin{overpic}
[width=1.1\textwidth]{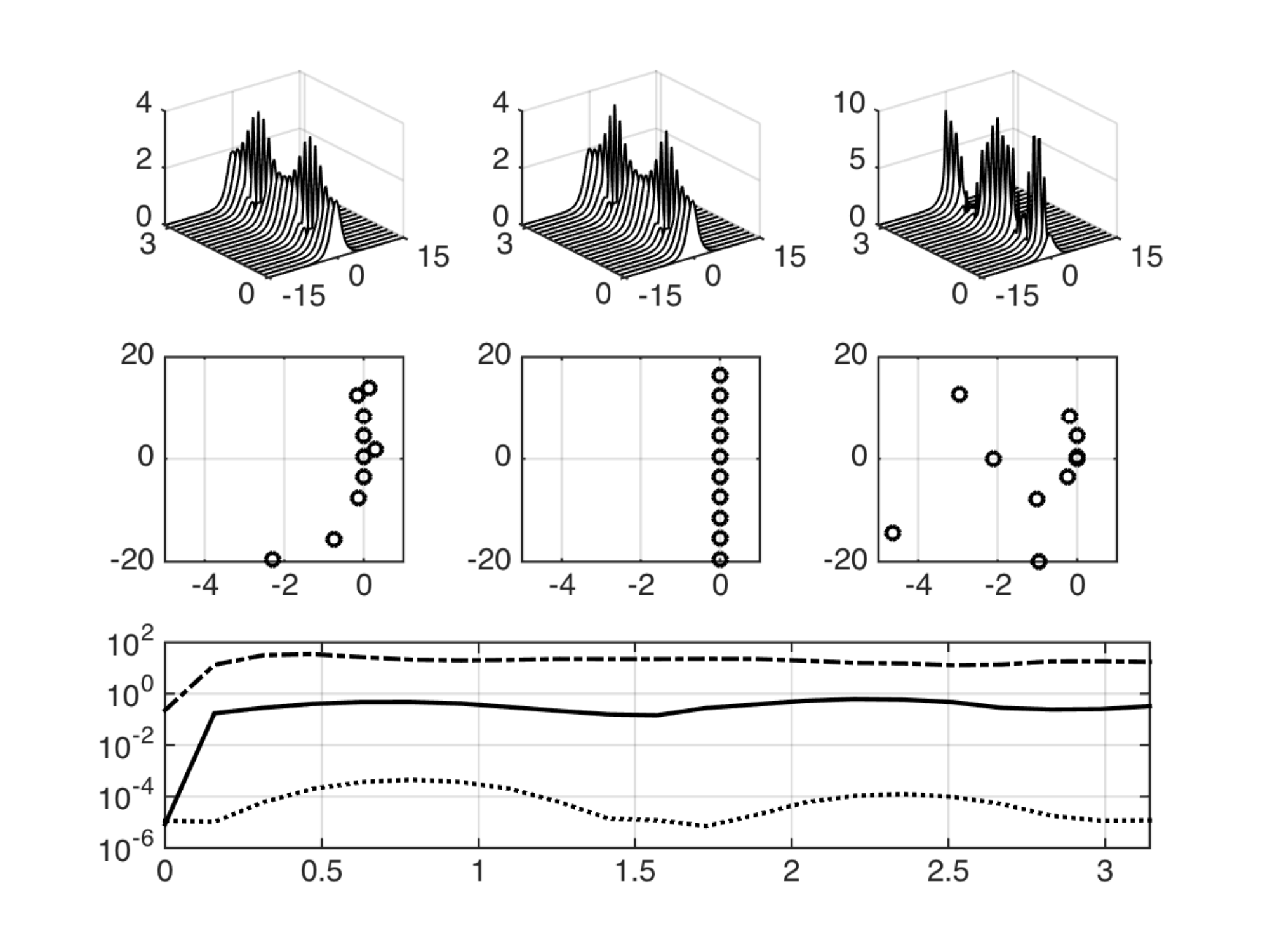}
\put(14,68){$(a)$}
\put(42,68){$(b)$}
\put(70,68){$(c)$}
\put(14,43){(d)}
\put(42,43){(e)}
\put(70,43){(f)}
\put(16,13){(g)}
\put(6,64){$|u|$}
\put(14,52){$t$}
\put(30,51){$x$}
\put(4,43){$\Re\{\omega_k\}$}
\put(24,26.5){$\Im\{\omega_k\}$}
\put(4,17){$E$}
\put(68,4){$t$}
\put(60,17){${\bf g}_{\mbox{\tiny DMD}} ({\bf x})$}
\put(65,10){${\bf g}_1({\bf x})$}
\put(52,21){${\bf g}_2({\bf x})$}
\end{overpic}
\vspace*{-0.25in}
\caption{Reconstruction of the NLS dynamics using (a) a standard DMD approximation ${\bf g}_{\mbox{\tiny DMD}} ({\bf x})$, 
(b) the NLS motivated ${\bf g}_1({\bf x})$ and (c) a quadratic observable ${\bf g}_2({\bf x})$.  The Koopman spectra
for each observable is demonstrated in panels (d), (e) and (f) which accompany the observables of (a), (b) and (c) respectively.  Note that
the observable ${\bf g}_1({\bf x})$ produces a spectra which is approximately purely imaginary which is expected of the 2-soliton evolution.
The error between the three observables and the full simulation is shown in panel (g).  Note that the observable ${\bf g}_1({\bf x})$
gives an error reduction of four orders of magnitude over DMD, while ${\bf g}_2({\bf x})$ is an order of magnitude worse.  This highlights
the importance of selecting good observables. 
\label{fig:nls1}}
\end{figure}
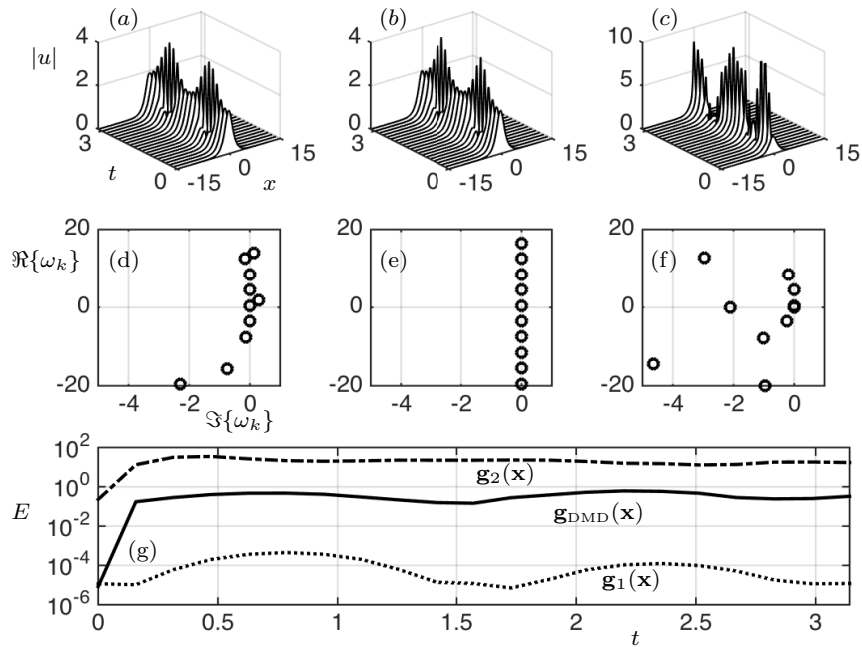

As with Burgers' equation, we can compute the DMD by collecting snapshots of the dynamics
over a specified time window.  Specifically, we consider simulations of the equation with initial data
\begin{equation}
  u(x,0)= 2\mbox{sech} (x)
  \label{eq:nlsi}
\end{equation}
over the time interval $t\in[0,\pi]$.  Twenty one snapshots of the dynamics are collected during the evolution, allowing
us to create the snapshot matrix ${\bf X}$ and ${\bf X}'$. The DMD reconstruction of the dynamics is demonstrated in Fig.~\ref{fig:nls1}(a).
The low-rank DMD reconstruction provides a good approximation to the dynamics of the PDE.

 To be more precise,  it is explicitly
assumed in the DMD reduction that the observables are simply the
state variables ${\bf x}$ where ${\bf x}=u(x,t)$ at discrete space and time points.  The DMD observables are then given by
\begin{equation}
  {\bf g}_{\mbox{\tiny DMD}} ({\bf x})= {\bf x} \, .
\end{equation}
Thus as previously noted,  the DMD approximation is a special case of Koopman.  The DMD
spectrum for a rank $r=10$ approximation is shown in Fig.~\ref{fig:nls1}(d).  An ideal approximation would
have the eigenvalues aligned along the imaginary axis since the evolution with the initial condition given by
(\ref{eq:nlsi}) is known as the 2-soliton solution which is purely oscillatory.

 Koopman theory allows us a much broader set of observables.  In what follows,
we consider two additional observables
\begin{subeqnarray}
  && {\bf g}_1({\bf x})=\left[  \begin{array}{cc} {\bf x}\\ |{\bf x}|^2 {\bf x} \end{array} \right]  \\
  && {\bf g}_2({\bf x})=\left[  \begin{array}{cc} {\bf x}\\ |{\bf x}|^2 \end{array}  \right] \, .
  \label{eq:observables}
\end{subeqnarray}
The first observable ${\bf g}_1({\bf x})$ is motivated by the form of the nonlinearity in the NLS equation.
The second, ${\bf g}_2({\bf x})$, is chosen to have a simple quadratic nonlinearity.  It has
no special relationship to the governing equations.  Note that the choice of the observable
$|{\bf x}|^2$ in ${\bf g}_2({\bf x})$ is relatively arbitrary.  For instance, one could consider
instead $|{\bf x}|^5 {\bf x}$, ${\bf x}^2$, ${\bf x}^3$ or
${\bf x}^5$, for instance.  These all produce similar results to the ${\bf g}_2({\bf x})$ selected
in (\ref{eq:observables}b).  Specifically, the observable ${\bf g}_2({\bf x})$ is inferior to
either the DMD or judiciously selected ${\bf g}_1({\bf x})$ for the Koopman reconstruction.

As has been repeatedly stated, the success of the Koopman decomposition relies almost exclusively on
the choice of observables.   To demonstrate this in practice, we compute the Koopman
decomposition of the NLS equation (\ref{eq:nls_koop}) using the two observables (\ref{eq:observables}).
The required data matrices have $2n$ rows of data, and only the state variables need to be
recovered at the end of the procedure.  Note that the algorithm produces both a state approximation 
since the first $n$ components are actually the state vector ${\bf x}$, as well as approximations to
the nonlinearity.   The Koopman eigenfunctions and eigenvalues provide information about the evolution
on the observable space.

Figure~\ref{fig:nls1}(b) and (d) show the Koopman reconstruction of the simulated data for the
observables (\ref{eq:observables}).   The observable ${\bf g}_1({\bf x})$ provides an exceptional 
approximation to the evolution while ${\bf g}_2({\bf x})$ is quite poor.  Indeed, the error of
the DMD approximation and two nonlinear observables (\ref{eq:observables})  
are shown Fig.~\ref{fig:nls1}(g) where the following error metric is used:
\begin{equation}
    E (t_k) = \| {\bf x}(t_k) - \tilde{\bf x}(t_k)   \|   \,\,\,\,\,\,\,\,\,\, k=1, 2, \cdots m
    \label{eq:koop_error}
\end{equation}
where ${\bf x}$ is the full simulation and $\tilde{\bf x}$ is the DMD or Koopman approximation.
With the choice of observable
${\bf g}_1({\bf x})$, which was judiciously chosen to match the nonlinearity of the NLS, the Koopman
approximation of the dynamics is four-orders of magnitude better than a DMD approximation.  A poor
choice of observables, given by ${\bf g}_2({\bf x})$, gives the worse performance of all, an order of magnitude
worse than DMD.  Note also
the difference in the Koopman spectral as shown in the middle panels of  Fig.~\ref{fig:nls1}.  In particular,
note that the judicious observables ${\bf g}_1({\bf x})$ aligns the eigenvalues along the
imaginary axis as is expected from the dynamics.  It further suggests that much better long-time predictions
can be achieved with the Koopman decomposition using ${\bf g}_1({\bf x})$.  

Observable selection in this case was facilitated by knowledge of the governing equations.  However, in many
cases, no such expert knowledge is available, and we must rely on data.  
The kernel DMD method allows one to use the kernel trick to consider a vast range of potential observables.
As already highlighted, the kernel method allows for an efficient method to consider a large class of potential observables
without making the observation vector ${\bf g}({\bf x})$ computationally intractable.
For instance, one can consider a radial basis function kernel
\begin{equation}
  f ({\bf x},{\bf x}')= \exp \left( - | {\bf x} - {\bf x}' |^2 \right) \, .
  \label{eq:rbf}
\end{equation}
The absolute value is important for the case of the NLS equation considered due to the nonlinear evolution of
the phase.   This radial basis function is one of the more commonly considered kernels.  
Other kernels that we might consider include the three following observables
\begin{subeqnarray}
  &&  f ({\bf x},{\bf x}')= \left( 1 + {\bf x}^T{\bf x}' \right)^{20}  \\
  &&  f ({\bf x},{\bf x}')= \left( a + | {\bf x}^T | |{\bf x}'| \right)^{20} \\
  &&  f ({\bf x},{\bf x}')=  \exp (- {\bf x}^T{\bf x}' ) \, .  \label{eq:kernel_choices}
\end{subeqnarray}
The first function is the standard polynomial kernel of 20th degree.   The second instead takes
the absolute value of the variable in a polynomial in order to remove the
phase and the third is a Gaussian kernel that uses the same inner product as
the polynomial kernel.   

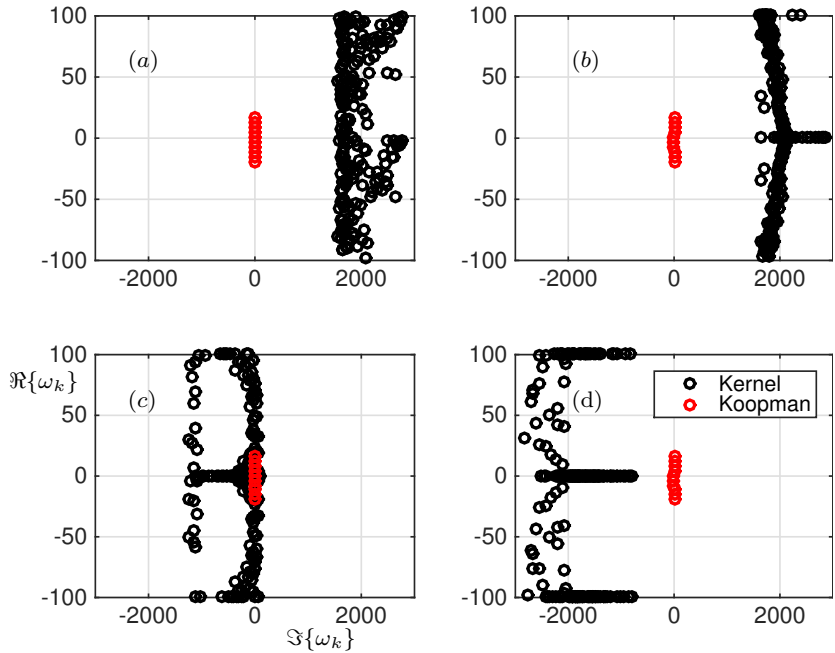
\begin{figure}[t]
\vspace*{.3in}
\hspace*{.20in}
\begin{overpic}
[width=0.9\textwidth]{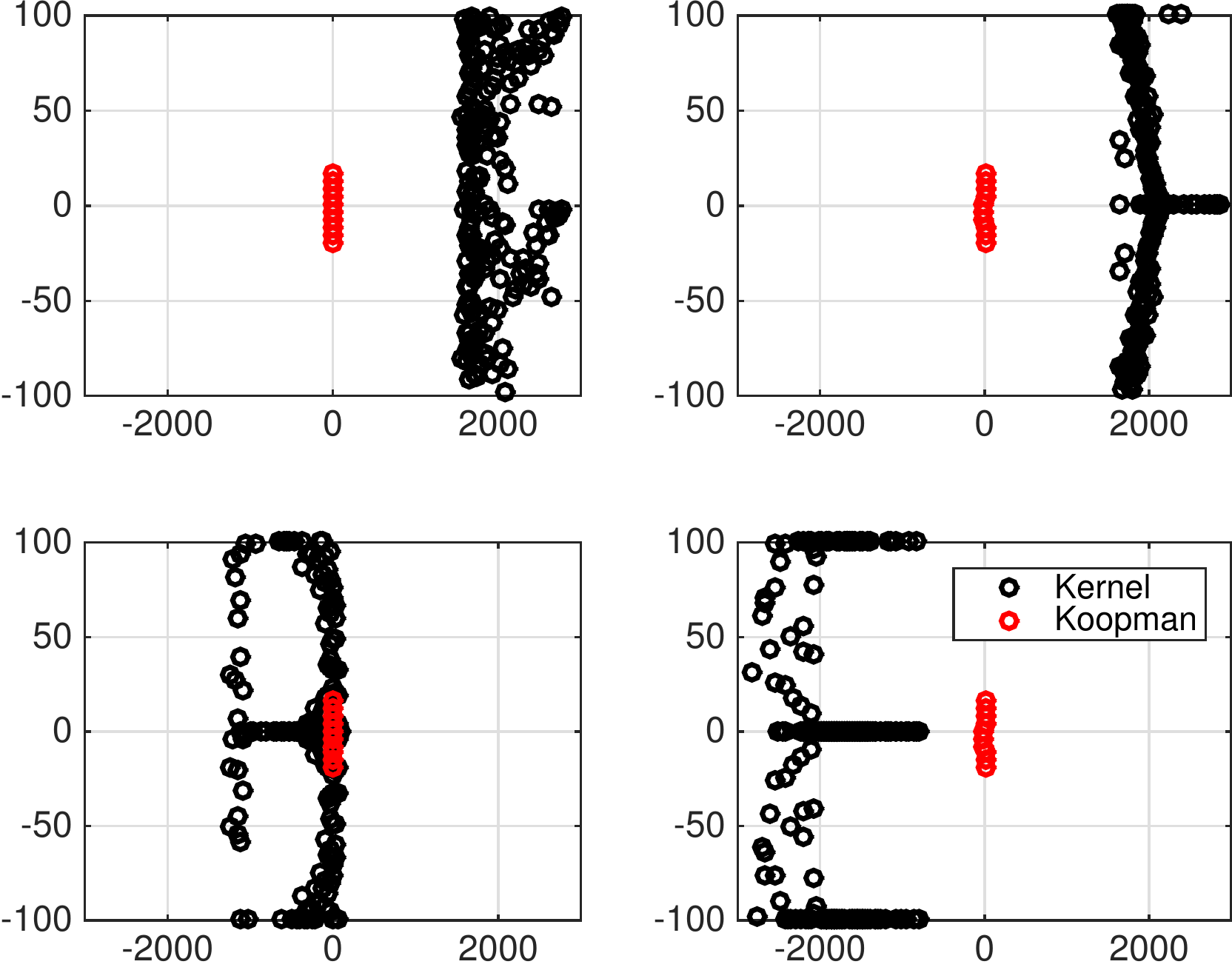}
\put(-4,30.5){$\Re\{\omega_k\}$}
\put(31,-2){$\Im\{\omega_k\}$}
\put(11,71){$(a)$}
\put(67,71){$(b)$}
\put(11,28){$(c)$}
\put(67,28){(d)}
\end{overpic}
\vspace*{0.1in}
\caption{Koopman spectra (a)-(d) of the four kernels considered in (\ref{eq:rbf}) and (\ref{eq:kernel_choices})(a)-(c) respectively.
The red spectra is the Koopman spectra generated from the rank $r=10$ observable ${\bf g}_1({\bf x})$ which provides an exceptionally
accurate reconstruction of the NLS dynamics.    
\label{fig:nls2}}
\end{figure}

These three new kernels are compared to each other and the radial
basis function.  Figure~\ref{fig:nls2} shows
the spectra generated by these four kernels along with a comparison to the
Koopman spectra generated by ${\bf g}_1({\bf x})$.  Note the tremendous variability of the
results based upon the choice of kernel.   Indeed, it highlights the tremendous sensitivity and
non-robust nature of the kernel method for selecting observables.  The choice of kernel must be carefully selected for either the extended or kernel DMD to give anything reasonable.  Cross validation techniques 
could potentially be used to select a  suitable kernel for applications of interest.  It could
also ensure that overfitting of the data does not occur.  In either case, this simple example should serve
as a strong cautionary tale for using kernel techniques in Koopman theory unless results are carefully cross validated.

\section{Outlook on Koopman Theory for PDEs}

Koopman analysis is a remarkable theoretical architecture with applicability to a wide range of nonlinear dynamical systems and PDEs.
It combines a number of innovations across disciplines, including dimensionality-reduction techniques, manifold learning, 
linear operator theory, and dynamical systems.  Although the abstract architecture provides a tremendously compelling viewpoint on
how to transform nonlinear dynamical systems to infinite-dimensional linear dynamics, significant challenges remain in
positing an appropriate set of observables for construction of the Koopman operator.   If good candidate observables can be found,
then the DMD algorithm can be enacted to compute a finite-dimensional approximation of the Koopman operator, including its
eigenfunctions, eigenvalues and Koopman modes.  With a judicious choice of observables, these computed quantities can often
lead to physically interpretable spatio-temporal features of the complex system under consideration.

We have demonstrated the application of Koopman theory on two canonical, nonlinear PDEs:  Burgers' equation and the nonlinear
Schr\"odinger equation.   For Burgers' equation, the well-known Cole-Hopf transformation provides a critical link to an explicit
calculation of the Koopman operator for a nonlinear PDE.  Indeed, we show that the Koopman operator and associated observables
can be trivially constructed from knowledge of the Cole-Hopf transformation.  In contrast, choosing linear state observables for
Burgers' yields a DMD approximation which is accurate, but lacks the clear physical interpretation of the exact Koopman reduction.
Although the NLS equation can similarly be linearized via the inverse scattering transform, the transform and its inverse are technically difficult to
compute for arbitrary initial conditions.  Instead, we demonstrate that the selection of an observable that is motivated by
the nonlinearity of the governing PDE gives a remarkably accurate Koopman reduction.  Indeed, the Koopman eigenfunctions
and eigenvalues provide and approximation that is nearly equivalent to the accuracy of the numerical simulation itself.  Importantly,
for the NLS example, we also demonstrate that poor choices of observables are significantly worse than the DMD approximation.
And for the case of observables chosen with a kernel method, the resulting spectra and eigenfunctions are highly inaccurate and
non-robust, suggesting that such generic techniques as kernel methods may face challenges for use in observable selection.

Ultimately, the selected observables do not need to exactly linearize the system, but they should provide a method for transforming 
a strongly nonlinear dynamical system to a weakly nonlinear dynamical system.  In practice, this is all that is necessary to make
the method viable and informative.  The results presented here are simultaneously compelling and concerning, highlighting 
the broader outlook of the
Koopman method in general.   Specifically, the success of the method will hinge on one issue:  selection of observables.
If principled techniques, from expert-in-the-loop knowledge, the form of the governing equation, or information about the
manifold on which the data exists, can be leveraged to construct suitable observables, then Koopman theory should
provide a transformative method for nonlinear dynamical systems and PDEs.  We posit that sparse statistical regression 
techniques from machine learning may provide a path forward towards achieving this goal of selecting quality observables~\cite{Brunton2016pnas,Brunton2016plosone}. 
Failing this, the Koopman architecture may have a limited impact in the mathematical sciences. 
Because of the importance of identifying meaningful observables, this is an exciting and growing area of research, especially given new developments in machine learning that may provide a robust and principled approach to observable selection.

\begin{acknowledgement}
J. N. Kutz would like to acknowledge support from the Air Force Office of Scientific Research (FA9550-15-1-0385). J.L. Proctor would like to thank Bill and Melinda Gates for their active support of the Institute for Disease Modeling and their sponsorship through the Global Good Fund.
\end{acknowledgement}
\bibliographystyle{unsrt}
\bibliography{bibALL}
\end{document}